\input harvmac.tex 

\newdimen\tableauside\tableauside=1.0ex
\newdimen\tableaurule\tableaurule=0.4pt
\newdimen\tableaustep
\def\phantomhrule#1{\hbox{\vbox to0pt{\hrule height\tableaurule width#1\vss}}}
\def\phantomvrule#1{\vbox{\hbox to0pt{\vrule width\tableaurule height#1\hss}}}
\def\sqr{\vbox{%
  \phantomhrule\tableaustep
  \hbox{\phantomvrule\tableaustep\kern\tableaustep\phantomvrule\tableaustep}%
  \hbox{\vbox{\phantomhrule\tableauside}\kern-\tableaurule}}}
\def\squares#1{\hbox{\count0=#1\noindent\loop\sqr
  \advance\count0 by-1 \ifnum\count0>0\repeat}}
\def\tableau#1{\vcenter{\offinterlineskip
  \tableaustep=\tableauside\advance\tableaustep by-\tableaurule
  \kern\normallineskip\hbox
    {\kern\normallineskip\vbox
      {\gettableau#1 0 }%
     \kern\normallineskip\kern\tableaurule}%
  \kern\normallineskip\kern\tableaurule}}
\def\gettableau#1 {\ifnum#1=0\let\next=\null\else
  \squares{#1}\let\next=\gettableau\fi\next}

\tableauside=1.0ex
\tableaurule=0.4pt

\def\IE{\relax{\rm I\kern-.18em E}}

\lref\cs{E. Witten, ``Quantum field theory and the Jones polynomial," 
Commun. Math. Phys. {\bf 121} (1989) 351.}
\lref\gv{R. Gopakumar and C. Vafa, ``On the gauge theory/geometry 
correspondence," hep-th/9811131, Adv. Theor. Math. Phys. {\bf 3} (1999) 1415.}
\lref\ov{H. Ooguri and C. Vafa, ``Knot invariants and topological 
strings," hep-th/9912123, Nucl. Phys. {\bf B 577} (2000) 419.} 
\lref\gvmone{R. Gopakumar and C. Vafa, ``M-theory and topological 
strings, I," hep-th/9809187. }
\lref\gvm{R. Gopakumar and C. Vafa, ``M-theory and topological 
strings, II," hep-th/9812127. }
\lref\kkv{S. Katz, A. Klemm and C. Vafa, ``M-theory, topological strings, and 
spinning black-holes," hep-th/9910181, 
Adv. Theor. Math. Phys. {\bf 3} (1999) 1445.}
\lref\fh{W. Fulton and J. Harris, {\it Representation theory. A first course},
Springer-Verlag, 1991.}
\lref\guada{E. Guadagnini, ``The universal link polynomial," 
Int. J. Mod. Phys. {\bf A 7} (1992) 877;  {\it The link invariants of the 
Chern-Simons field theory,} Walter de Gruyter, 1993.}
\lref\lick{W.B.R. Lickorish, {\it An introduction to knot theory}, 
Springer-Verlag, 1998.}
\lref\ofer{O. Aharony, S. Gubser, J. Maldacena, H. Ooguri and Y. Oz, 
``Large $N$ field theories, string theory and gravity," hep-th/9905111, 
Phys. Rep. {\bf 323} (2000) 183.}
\lref\lpp{J.M.F. Labastida and E. P\'erez, ``Gauge-invariant 
operators for singular knots in
Chern-Simons gauge theory," hep-th/9712139, Nucl.\ Phys.\ {\bf
B 527} (1998) 499.}
\lref\witop{E. Witten, ``Chern-Simons gauge theory as 
a string theory,'' hep-th/9207094, in {\it The Floer memorial volume}, 
H. Hofer, C.H. Taubes, A. Weinstein and E. Zehner, eds., 
Birkh\"auser 1995, p. 637.}  
\lref\lm{J.M.F. Labastida and M. Mari\~no, ``Polynomial invariants 
for torus knots and topological strings,''  hep-th/0004196,
Commun. Math. Phys. {\bf 217} (2001) 423.}
\lref\lmv{J.M.F. Labastida, M. Mari\~no and C. Vafa, ``Knots, links and 
branes at large $N$,'' hep-th/0010102, JHEP {\bf 0011} (2000) 007.}
\lref\kl{S. Katz and M. Liu, ``Enumerative geometry of stable 
maps with Lagrangian boundary conditions and multiple covers of the disc,'' 
math.AG/0103074.}
\lref\k{M. Kontsevich, ``Intersection theory on the moduli space of 
curves and the matrix Airy function," Commun. Math. Phys. {\bf 147} (1992) 1.} 
\lref\ls{J. Li and Y.S. Song, ``Open string instantons and relative 
stable morphisms,'' hep-th/0103100.}
\lref\rama{P. Ramadevi and T. Sarkar, ``On link invariants and 
topological string amplitudes,'' hep-th/0009188, Nucl. Phys. {\bf B 600} 
(2001) 487.}
\lref\sv{S. Sinha and C. Vafa, ``SO and Sp Chern-Simons at large $N$,'' 
hep-th/0012136.}
\lref\gmm{E. Guadagnini, M. Martellini and M. Mintchev, ``Wilson 
lines in Chern-Simons theory and link invariants,'' Nucl. Phys. 
{\bf B 330} (1990) 575.}
\lref\newpoint{J.M.F. Labastida and M. Mari\~no, ``A new point of 
view in the theory of knot and link invariants,'' math.QA/0104180.}
\lref\polyakov{A.M. Polyakov, ``Fermi-Bose transmutations induced by 
gauge fields,'' Mod. Phys. Lett. {\bf A 3} (1988) 325.}
\lref\av{M. Aganagic and C. Vafa, ``Mirror symmetry, D-branes, and counting 
holomorphic discs,'' hep-th/0012041.}
\lref\akv{M. Aganagic, A. Klemm and C. Vafa, ``Disk instantons, mirror 
symmetry, and the duality web,'' hep-th/0105045.}
\lref\wtwodg{E. Witten,  
``Two-dimensional gravity and 
intersection theory on the moduli space," Surveys in Differential Geometry 
{\bf 1} (1991) 243.} 
\lref\stru{E. Witten, ``On the structure of the topological phase 
of two-dimensional gravity," Nucl. Phys. {\bf B 340} (1990) 281.}
\lref\hm{J. Harris and I. Morrison, {\it Moduli of curves}, Springer-Verlag, 
1998.}
\lref\faber{C. Faber, ``Algorithms for computing intersection numbers 
of curves, with an application to the class of the locus of Jacobians," 
alg-geom/9706006.}
\lref\hint{E. Getzler and R. Pandharipande, ``Virasoro constraints and 
the Chern classes of the Hodge bundle," math.AG/9805114, Nucl. Phys. 
{\bf B 530} [PM] (1998) 701. C. Faber and R. Pandharipande, ``Hodge 
integrals and Gromov-Witten theory," math.AG/9810173, Invent. Math. {\bf 139} 
(2000) 173. ``Logarithmic series and Hodge integrals in the 
tautological ring," math.AG/0002112.}
\lref\mumford{D. Mumford, ``Towards an enumerative geometry of the moduli 
space of curves,'' in {\it Arithmetic and Geometry}, Vol. II, p. 271, 
Birkh\"auser 1983.}

\Title{\vbox{\baselineskip12pt
\hbox{RUNHETC-2001-23}
\hbox{HUTP-01/A039}
\hbox{hep-th/0108064}
}}
{\vbox{\centerline{Framed Knots at Large $N$}
}}
\centerline{Marcos Mari\~no$^{a}$ and Cumrun Vafa$^{b}$}

\bigskip
\medskip
{\vbox{\centerline{$^{a}$ \sl New High Energy Theory Center, 
Rutgers University}
\vskip2pt
\centerline{\sl Piscataway, NJ 08855, USA}}

\bigskip
\medskip
{\vbox{\centerline{$^{b}$ \sl Jefferson Physical Laboratory, 
Harvard University}
\vskip2pt
\centerline{\sl Cambridge, MA 02138, USA }}}

\bigskip
\bigskip
\noindent

We study the framing dependence
of the Wilson loop observable of $U(N)$
Chern-Simons gauge theory at large $N$.  Using proposed
geometrical large $N$ dual, this leads to a direct computation of
certain topological string amplitudes in a closed form.  This
yields new formulae for intersection numbers of cohomology
classes on moduli of Riemann surfaces with punctures
(including all the amplitudes of pure topological
gravity in two dimensions).  The reinterpretation
of these computations in terms of BPS degeneracies of domain walls leads to
novel integrality predictions for these amplitudes.
Moreover we find evidence that large $N$ dualities 
are more naturally formulated in the context of $U(N)$ gauge
theories rather than $SU(N)$.

\bigskip

\Date{August, 2001}

\newsec{Introduction}
Topological strings have been studied for about a decade now.  Most
of the progress in their study has been in the context of closed strings
(i.e. Riemann surfaces without boundaries).  In the type II superstrings
these amplitudes correspond to  certain computation for the effective
4 dimensional theory with ${\cal N}=2$ supersymmetry (or more generally
for theories with 8 supercharges).
More recently
progress has been made also in understanding topological strings involving
open strings, i.e. with D-branes in target space geometry.  These
translate in the context of superstrings to F-term computations
for the underlying 4 dimensional theory with ${\cal N}=1$ (or more
generally for theories with 4 supercharges).

This progress has come from three different directions:  On the one
hand large $N$ duality of Chern-Simons gauge theory 
on ${\bf S}^3$ with closed topological strings on ${\cal O}(-1)\oplus
{\cal O}(-1) \rightarrow {\bf P}^1$ \gv\ in the context of
Wilson loop observables at large $N$ formulated in \ov\
has led to predictions for a special class of open topological
string amplitudes for all genera and arbitrary number of holes
\ov\lm\rama\lmv\newpoint .  On the other hand
techniques from mirror symmetry have led to computations
of disc amplitudes for a large class of open string topological
amplitudes \av\akv .  Finally,  more recently, from a direct mathematical
computation of Gromov-Witten invariants using localization ideas
a number of results have emerged \kl\ls\ref\zas{
T. Graber and E. Zaslow, in preparation.}\ that lead in principle to computation
of the amplitudes at all genera and arbitrary number
of holes for a large class of open topological
amplitudes, in terms of intersection theory on moduli of 
Riemann surfaces with punctures, for which there are known
computational algorithms.

However, it was discovered in \akv , that there is an inherent ambiguity
in the open topological string amplitude related to the IR
geometry of the D-brane.   Moreover it was shown 
there that in the context of duality with large $N$ Chern-Simons theory,
this gets mapped to the well known framing ambiguity for knot invariants
\cs .  The existence of this choice, labeled by an integer, was
also verified in \kl\zas .  

A major goal of this paper is to extend the computation of \akv\
for the framing dependence of the amplitudes, which
was carried out for disc amplitudes, to arbitrary genus Riemann surfaces with 
holes, for D-branes which are large $N$ duals of knots.
For the case of the unknot in a special limit (the limit
of large ${\bf P}^1$ volume) the framing dependence of Gromov-Witten
invariants was computed for all Riemann surfaces with boundaries, where
it was reduced to certain intersection classes on moduli of Riemmann
surfaces with punctures, which can be computed
using known algorithms \faber\hint.  Given our result we find
a closed expression for all such intersections in terms of framing
dependence of the unknot. This also turns out to exhibit certain
novel integrality properties.

The organization of this paper is as follows:  In section 2
we review aspects of topological strings in the context of Riemann
surfaces with boundaries.  This includes the discussion about
integrality predictions for these amplitudes.  In section 3
we review the framing dependence of 
knots for Chern-Simons theory.  In section 4 we check that the large
$N$ expansion for knots for Chern-Simons theory has the expansion 
and integrality
properties expected for open topological strings
for arbitrary framing.  In section 5 we compare
the special case of our results to the results of \kl\ which leads
to a closed formula for intersection theory of Mumford classes
with insertion of up to three Chern classes of the Hodge bundle
on moduli of punctured Riemann surfaces. In section
6 we compute using techniques from mirror
symmetry \av \akv\ the framing
dependence of the unknot for arbitrary volume for ${\bf P}^1$ and
find agreement with the framing dependence of the large $N$ limit of unknot.

\newsec{Open topological string}

Open topological string computes certain invariants
related to the space of holomorphic maps from Riemann surfaces with
boundaries
to Calabi-Yau manifolds where the boundary lies on a Lagrangian submanifold,
identified with a topological D-brane.
These invariants are called Gromov-Witten invariants.  For simplicity,
and in view of the application to the large $N$ dual of Chern-Simons
theory, consider the case where the Calabi-Yau manifold $X$ has one K\"ahler
moduli denoted by $t$ and assume that the Lagrangian submanifold ${\cal C}$
has one non-trivial $H_1$. Let us assume we wrap $M$ D-branes
around ${\cal C}$ and denote the holonomy around the non-trivial
$H_1$ by the matrix $V$. 

Let us consider the topological string theory associated 
to the maps of an open Riemann surface $\Sigma_{g,h}$ (with genus $g$ and 
$h$ holes) to $X$ with holes mapped to ${\cal C}$.  
To each boundary we can associate an integer
related to how many times it wraps the corresponding element
of ${\cal C}$: if the $i$-th 
hole winds around the one-cycle $n_i$ 
times, the homotopy class of the boundary can be labeled by $h$ integers 
$\vec n =(n_1, \cdots, n_h)$. The free energy of topological string 
theory in the topological sector 
labeled by $\vec n$ can be regarded as a generating 
functional of open Gromov-Witten 
invariants:
\eqn\freen{
F_{g, \vec n} (t) =\sum_Q F_{\vec n, g}^{Q} {\rm e}^{-Q t}.}
In this equation, $t$ is the complexified K\"ahler parameter of the 
Calabi-Yau manifold, and $Q$ labels the relative homology class of  
the embedded Riemann surface. 
The quantities $F_{g, \vec n}^{Q}$ 
are the open string analog of the Gromov-Witten invariants 
and they ``count" in an appropriate sense the number of holomorphically 
embedded Riemann surfaces of genus $g$ in $X$ with Lagrangian boundary 
conditions specified by ${\cal C}$ with the class represented
by $Q,{\vec n}$.  These are in general rational numbers.
Mathematical aspects of defining these quantities
have been considered recently in \kl\ls\zas. 

We can now consider the total free energy, which is 
the generating functional for all topological 
sectors:
\eqn\totaln{
F(V)=\sum_{g=0}^{\infty} \sum_{h=1}^{\infty} 
\sum_{n_1, \cdots, n_h} {i^h \over h!}
g_s^{2g-2+h} F_{g, \vec n} (t) 
{\rm Tr}\,V^{n_1} \cdots {\rm Tr}\, V^{n_h},}
where $g_s$ is the string coupling constant. 
The factor $i^h$ is very convenient 
in order to compare to the Chern-Simons free energy, as we will see 
in a moment. The factor $h!$ is a symmetry factor which takes into 
account that the holes are indistinguishable (or one could have
absorbed them into the definition of $F_{g,{\vec n}}$).  We take
all $n_i >0$ (as discussed in \ov\ this can be achieved if necessary
by analytic continuation of the amplitude).

It is convenient to rewrite \totaln\ 
in terms of a vector $\vec k$. Given a vector $\vec n=(n_1, \cdots, n_h)$, 
we define a vector $\vec k$ as follows: the $i$-th 
entry of $\vec k$ is the number of $n_j$'s 
which take the value $i$. For example, if $n_1=n_2=1$ and 
$n_3=2$, then this corresponds to $\vec k =(2,1,0,\cdots)$. 
In terms of $\vec k$, the number of holes and the total winding number are  
given by 
\eqn\quantis{
h=|\vec k|\equiv \sum_J k_j,\,\,\,\ \ell =\sum_i n_i=\sum_j  j k_j.}
Note that a given $\vec k$ will correspond to many ${\vec n}$'s
which differ by permutation of entries.  In fact
there are $h!/\prod_j k_j!$ vectors $\vec n$ which 
give the same vector $\vec k$ (and the same amplitude). We can 
then write the total free energy as:  
\eqn\totaln{
F(V)=\sum_{g=0}^{\infty} \sum_{\vec k} {i^{|\vec k|} \over \prod_j k_j!}
g_s^{2g-2+h} F_{g, \vec k} (t) \Upsilon_{\vec k} (V)}
where
$$\Upsilon_{\vec k} (V)=\prod_{j=1}^{\infty} ({\rm Tr} V^j)^{k_j}$$
%

\subsec{Integrality properties}

Let us define 
$$q=e^{ig_s},\qquad \lambda =e^t.$$
We now define the generating functions $f_R (q, \lambda)$ through the 
following equation:
 \eqn\conj{
F(V)= -\sum_{n=1}^\infty \sum_{R} 
{1\over n} f_R (q^n, \lambda^n) {\rm Tr}_R 
{V^n} } 
where $R$ denotes a representation of $U(M)$ and we are considering
the limit $M\rightarrow \infty$. In this limit we can exchange the basis
consisting
of product of traces of powers in the fundamental representation, with
the trace in arbitrary representations.
It was shown in \ov , following
similar ideas in the closed string case \gvm , that the open topological
strings compute the partition function of BPS domain walls in a related
superstring theory.  This led to the result that $F(V)$ has an integral
expansion structure. This result was further refined in \lmv\
where it was shown that the corresponding integral expansion
leads to the following formula for
 $f_R (q, \lambda)$:
\eqn\fr{
f_{R}(q, \lambda)=
\sum_{g\ge 0} \sum_{Q}
\sum_{R', R''} 
C_{R\,R'\,R''}S_{R'}(q) 
{\widehat N}_{R'',g,Q} 
 (q^{-{1\over 2}}-q^{1\over 2})^{2g-1}\lambda^Q.}
In this formula $R,R',R''$ label representations of the symmetric group 
$S_\ell$, which can be labeled
by a Young tableau with a total of $\ell$ boxes.
In this equation, $C_{R\,R'\,R''}$ are the Clebsch-Gordon coefficients
of the symmetric group, and the monomials $S_R (q)$ are defined as 
follows. If $R$ is a hook representation 
\eqn\hook{
\tableau{6 1 1 1 1}} 
with $\ell$ boxes in total, and with $\ell-d$ boxes in the first row, then 
\eqn\expsr{
S_R (q)=(-1)^d q^{ -{\ell -1 \over 2}+d} ,} 
and it is zero otherwise. The ${\widehat N}_{R,g,Q}$ are 
{\it integers} and compute the net number
of BPS domain walls of charge $Q$ and spin $g$ transforming
in the representation $R$ of $U(M)$, where we are using
the fact that representations of $U(M)$ can also be labeled by Young
tableaux. 

\subsec{Multicovering formulae}

In order to exhibit the multicovering aspects
of $F(V)$, it is convenient to use the 
related invariants \lmv\
\eqn\nminus{
n_{\vec k, g, Q} = \sum_R \chi_R (C(\vec k)) {\widehat N}_{R,g,Q},}
where $\chi_R (C(\vec k))$ are the characters of the symmetric group,
and $C(\vec k)$ denotes the conjugacy class of the symmetric group
with $k_i$ cycles of size $i$. 
Notice that these invariants are not as fundamental as ${\widehat N}_{R,g,Q}$. 
For example, integrality of $n_{\vec k, g, Q}$ follows from integrality 
of ${\widehat N}_{R,g,Q}$ (since the characters are integers), 
but not the other way around, and there are some further
integrality constraints on $n_{\vec k, g, Q}$.

The multicovering formula derived in \lmv\ states that the free energies 
of open topological string theory with a fixed homotopy class 
$\vec k$ can be written in terms 
of the integer invariants 
$n_{\vec k, g, Q}$ as follows:
\eqn\multicov{
\eqalign{
& \sum_{g=0}^{\infty} 
g_s^{2g-2 +|\vec k|} 
F_{g, \vec k}(t) = \cr &
\,  {1\over \prod_j j^{k_j}} 
\sum_{d|\vec k}  (-1)^{|\vec k|+g}\,
n_{ \vec k_{1/d}, g, Q}\, d^{|\vec k|-1}
\biggl( 2\sin {d g_s \over 2} \biggr)^{2g-2} 
\prod_j \biggl( 2\sin {j g_s \over 2} \biggr)^{k_j} \lambda^{Qd}.\cr}}
Notice there is one such identity for each $\vec k$.
In this expression, the sum is over all integers $d$ which satisfy the 
following condition: $d|j$ for every $j$ with $k_j \not= 0$. 
When this is the case, we define the vector $\vec k_{1/d}$ 
whose components are 
$(\vec k_{1/d})_i =\vec k_{di}$. Remember that $|\vec k|=h$ is the number 
of holes.
 
As shown in \lmv, the formula \multicov\ is in fact a 
consequence of \fr , and gives a 
natural generalization of 
the expression derived in \gvm\ for the 
closed string case. Just like the multicovering formula of 
\gvm\ expresses the usual Gromov-Witten invariants in terms of 
other integer invariants, the formula \multicov\ implies that 
the open Gromov-Witten invariants $F_{\vec k,g}^Q$ can be 
also written in terms of the integer invariants \nminus. Up to 
genus $2$ one finds, 
\eqn\expl{
\eqalign{
F_{\vec k, g=0}^{Q}=& (-1)^{|\vec k|} 
\sum_{d|\vec k} d^{|\vec k|-3} n_{ \vec k_{1/d} , 0, Q/d} ,\cr
F_{\vec k ,g=1}^{Q}=&-(-1)^{|\vec k|} 
\sum_{d|\vec k} \biggl( d^{|\vec k| -1} n_{\vec k_{1/d},1, Q/d}- 
{d^{|\vec k| -3} \over 24}\bigl(2d^2-\sum_j j^2k_j\bigr)\, 
n_{\vec k_{1/d},0, Q/d}\biggr) ,\cr 
F_{\vec k, g=2}^{Q}=&(-1)^{|\vec k|}
\sum_{d|\vec k} \biggl(d^{ |\vec k|+1}   n_{\vec k_{1/d}, 2, Q/d} + 
{d^{ |\vec k|-1}   \over 24 }n_{\vec k_{1/d}, 1, Q/d} \sum_j j^2 k_j 
 \cr & 
+ {d^{|\vec k| - 3} \over 5760} 
\bigl(24 d^4 - 20 d^2 \sum_j j^2 k_j -2\sum_j j^4k_j + 
5\sum_{j_1, j_2}j_1^2 j_2^2 k_{j_1} k_{j_2} \bigr)\, 
n_{\vec k_{1/d}, 0, Q/d} \biggr) .\cr}}
In these equations, the integer $d$ has to divide the vector $\vec k$ (in the 
sense explained above) and it is understood that $n_{\vec k_{1/d},g,Q/d}$ 
is zero if $Q/d$ is not a relative homology
class.

\newsec{$U(N)$ Chern-Simons theory and the framed knots}

The duality between Chern-Simons theory on ${\bf S}^3$ and topological 
string theory on the resolved conifold was formulated in \gv\ for the $SU(N)$ 
gauge theory. At the level of partition function, which was
checked in \gv, the difference between the $U(N)$
and $SU(N)$ Chern-Simons gauge theories
is an additive constant which is not unambiguously defined
in the context of topological strings.  However,
as we will see in this and the next section,
consideration 
of Wilson loop observables indicates that the duality is in fact far more
natural for  
the $U(N)$ theory, especially when one takes
into account the framing dependence, as we will discuss below.
This also suggests that perhaps also in other large $N$ superstring
duals, it is the $U(N)$ gauge theory which is dual to the
string theory.  In fact $U(N)$ gauge theory is 
the more natural version of the large
$N$ duality that `t Hooft proposed.

\subsec{Wilson loops in $U(N)$ Chern-Simons}

Chern-Simons theory has an action given by
\eqn\csaction{
S(A)= {k \over 4\pi} \int_M {\rm Tr} \Bigl( A\wedge d A + {2 \over 3} A
\wedge A \wedge A \Bigr),} 
where $A$ is a gauge connection for a gauge group $G$. The generators 
of the Lie algebra of $G$, $T^a$, are normalized as ${\rm Tr} (T^a T^b) =
-\delta^{ab}$. The gauge-invariant observables of this theory 
are the Wilson loop operators, which are defined as 
follows. Let $U$ be the holonomy of the gauge connection $A$ 
around a knot ${\cal K}$ described 
by a loop $\gamma$
\eqn\holo{
U_{\gamma}={\rm P}\,\exp\, \oint_{\gamma} A.} The Wilson loop operator 
in the representation $R$ is
\eqn\wilson{
{\rm Tr}_R U_{\gamma} .}
We will denote the vevs of products of Wilson loops by
\eqn\denote{
W_{(R_1, \cdots, R_L)}=\langle \prod_i {\rm Tr}_{R_i} U_{\gamma_i} \rangle,}
where the vev is a normalized one ({\it i.e.} we divide by $Z({\bf S}^3)$).

If we take $G=U(1)$, Chern-Simons theory turns out to be extremely simple, 
since it is essentially a Gaussian theory \polyakov.  The different 
representations are labeled by integers, and in particular the vevs 
of Wilson loop operators
\eqn\vevw{
\langle \prod_i \exp \bigl( n_i \int_{\gamma_i} A \bigr) \rangle }
can be computed exactly. In order to compute them, however, one has to 
choose a framing for each of the knots $\gamma_i$. This arises as follows: 
in evaluating the vev, contractions of the holonomies 
corresponding to different $\gamma_i$ produce the following integral:
\eqn\linking{
{\rm lk} ({\cal K}_i, {\cal K}_j)=
{1 \over 4 \pi} \oint_{\gamma_i}dx^{\mu} \oint_{\gamma_j} dy^{\nu} 
\epsilon_{\mu \nu \rho} { (x-y)^{\rho} \over |x-y|^3}.}
This is in fact a topological invariant, {\it i.e.} it is invariant 
under deformations of the contours $\gamma_i$, $\gamma_j$ and it is in fact 
the linking number of the knots ${\cal K}_i$ and ${\cal K}_j$.   
On the other hand, contractions of the holonomies corresponding to the 
same knot $\gamma$ involve the integral
\eqn\cotor{
\phi ({\cal K})={1 \over 4 \pi} \oint_{\gamma}dx^{\mu} \oint_{\gamma} dy^{\nu} 
\epsilon_{\mu \nu \rho} { (x-y)^{\rho} \over |x-y|^3}.}
This integral is well-defined and finite (see, for example, \gmm\guada), and it is 
called the cotorsion of $\gamma$. The problem is that the cotorsion 
is not invariant under deformations of the knot. In order to 
preserve topological invariance one has to choose another 
definition of the composite operator $(\int_{\gamma}A)^2$ by means of a 
framing. A framing of the knot consists of 
choosing a contour $\gamma^f$ for $\gamma$, specified by a normal vector 
field $n$. The cotorsion $\phi({\cal K})$ becomes then 
\eqn\regul{
\phi_f ({\cal K})={1 \over 4 \pi} 
\oint_{\gamma}dx^{\mu} \oint_{\gamma^f} dy^{\nu} 
\epsilon_{\mu \nu \rho} { (x-y)^{\rho} \over |x-y|^3} =
{\rm lk} ({\cal K}, {\cal K}^f).}
The correlation function that we obtain in this way is 
a topological invariant (a linking number) but the 
price that we have to pay is that our regularization depends on a set 
of integers $p_i ={\rm lk} ({\cal K}_i, {\cal K}^f_i)$ (one for each knot). 
The vev \vevw\ can now be computed, after choosing the framings, as 
follows:
\eqn\vevans{
\langle \prod_i \exp \bigl( n_i \int_{\gamma_i} A \bigr) \rangle = 
\exp \biggl( { \pi i \over k} \sum_i n_i^2 p_i + {\pi i \over k} 
\sum_{i \not= j} n_i n_j {\rm lk} ({\cal K}_i, {\cal K}_j) \biggr).}
This regularization is nothing but the `point-splitting' method
familiar in the context of QFT's.

Let us now consider a $U(N)$ Chern-Simons theory. The $U(1)$ factor 
decouples from the $SU(N)$ theory, and all the vevs factorize into an 
$U(1)$ and an $SU(N)$ piece. A representation of $U(N)$ is labeled 
by a Young tableau, and it decomposes into a representation of $SU(N)$ 
corresponding to that tableau, and a representation of $U(1)$ with charge:
\eqn\chargeuone{
n ={\ell \over {\sqrt N}},}
where $\ell$ is the number of boxes in the Young tableau. In order 
to compute the vevs associated to the $U(1)$ of $U(N)$, one has to take also into 
account that the coupling constant $k$ is shifted as $k \rightarrow k+N$. 
We then find that the vev of a product of $U(N)$ Wilson loops in 
representations $R_i$ is given by:
\eqn\factoriz{
W_{(R_1, \cdots, R_L)}^{U(N)} = 
\exp \biggl( { \pi i \over N(k+ N) } \sum_i \ell_i^2 p_i + 
{\pi i \over N(k +N)} 
\sum_{i \not= j} \ell_i \ell_j {\rm lk} ({\cal K}_i, {\cal K}_j) \biggr) 
W_{(R_1, \cdots, R_L)}^{SU(N)},} 
where the $SU(N)$ vev is computed in the framing specified by $p_i$. Notice 
that, in the case of knots, the $SU(N)$ and 
$U(N)$ computations differ in a factor which only depends on the 
choice of framing, while for links the answers also differ in a 
topological piece involving the linking numbers.  
For knots and links in ${\bf S}^3$, there is a standard or canonical 
framing, defined by asking that the self-linking number is zero. This   
corresponds to $p_i=0$, and in that case we find:
\eqn\sf{
W_{(R_1, \cdots, R_L)}^{U(N), \, {\rm sf}} = 
\exp \biggl( {\pi i \over N(k +N)} 
\sum_{i \not= j} \ell_i \ell_j {\rm lk} ({\cal K}_i, {\cal K}_j) \biggr) 
W_{(R_1, \cdots, R_L)}^{SU(N),\, {\rm sf}}.}
This is precisely the corrected vev that was introduced in \lmv, eq. (4.40), 
in order to match the Chern-Simons vevs with the topological string 
answer. This indicates that the geometric duality advocated in \gv\ is 
in fact a duality between $U(N)$ Chern-Simons gauge theory in ${\bf S}^3$ 
and topological string theory in the resolved conifold. We will find more 
evidence for this when we analyze the framing dependence.

\subsec{Framing dependence}

One can now study the effect of a change of framing on the 
vacuum expectation values of Wilson loops, in a general Chern-Simons 
theory with gauge group $G$. Consider $L$ Wilson loops $W_{R_i}$ in 
representations $R_i$ of $G$, with $i=1, \cdots, L$. 
It was shown in \cs\ that, under a change 
of framing of ${\cal K}_i$ by $p_i$ units, the vev of the product of 
Wilson loops changes as follows:
\eqn\naframing{
W_{(R_1, \cdots, R_L)}  \rightarrow \exp \biggl[ 2\pi i \sum_i 
p_i  h_{R_i} \biggr]W_{(R_1, \cdots, R_L)}
,} 
In this equation, $h_R$ is 
the conformal weight of the WZW primary field corresponding 
to the representation $R$, and it is given by:
\eqn\cweight{
h_R = {C_R\over 2(k+N)},}
where $C_R$ is the quadratic Casimir of the group $G$ in the 
representation $R$. When $G=U(N)$, the representations $R$ can 
be labeled by the lengths of rows in a Young tableau, $l_i$, 
where $l_1 \ge l_2 \ge \cdots$. In terms of these, the 
quadratic Casimir for $U(N)$ reads,
\eqn\explcas{
C_R = N \ell + \kappa_R,}
where $\ell$ is the total number of boxes in the tableau, and 
\eqn\kapar{
\kappa_R =\ell +  \sum_i \bigl( l_i^2 -2il_i  \bigr).}
For $SU(N)$, one has 
\eqn\explcasun{
C_R^{SU(N)} = N \ell + \kappa_R -{\ell^2 \over N}.}
Notice that the difference between the change of $SU(N)$ and 
$U(N)$ vevs under the change of framing is consistent with \factoriz. 
If we now introduce the variables (in anticipation of the large $N$ duality \gv )
\eqn\varias{
q=\exp \biggl( {2\pi i \over k+N}\biggr), \,\,\,\,\ \lambda=q^N,}
we see that $U(N)$ vevs change, under the change of 
framing, as
\eqn\unframing{
W_{(R_1, \cdots, R_L)} \rightarrow  q^{{1 \over 2}\sum_i
 \kappa_{R_i} p_i } \lambda^{{1\over 2} 
\sum_i \ell_i p_i} W_{(R_1, \cdots, R_L)}.}

\newsec{Framing dependence and the large $N$ dual}

It was proposed in \gv\ that large $N$ limit of
Chern-Simons gauge theory is given by closed
topological strings on ${\cal O}(-1) \oplus {\cal O}(-1)\rightarrow {\bf P}^1$,
with the identification of parameters $q,\lambda$ as given in sections 2 and 3.
This was formulated in the context of the string
theory realization of Chern-Simons theory on $T^*S^3$ with
$N$ Dbranes wrapping $S^3$ \witop.
In the large $N$ duality proposed in \gv\ the conifold undergoes
a geometric transition to the resolved conifold ${\cal O}(-1)\oplus
{\cal O}(-1)\rightarrow
{\bf P}^1$ where the branes wrapping $S^3$ have disappeared.
In the context of the large $N$ duality
 the formulation of Wilson loop observables
was studied in \ov\ which we now review.

\subsec{Review of the large $N$ duality for Wilson loops}

For simplicity let us consider the case of knots, as the extension
to links is straightforward.
The formulation of Wilson loop observables was achieved in \ov\
by considering $M$ branes in the $T^*S^3$ geometry where
$M$ branes wrap a Lagrangian submanifold intersecting
$S^3$ on the knot.
Moreover it was conjectured that for each knot in the large
$N$ dual, the $M$ branes deform to a Lagrangian submanifold
in the resolved conifold geometry.  This was shown for algebraic
knots in \lmv .  More recently it has been extended to all knots
\ref\taub{C. Taubes, to appear.}.
Moreover, the partition function of open topological string $F(V)$
with $M$ branes in the resolved
conifold geometry is related to the Wilson loop observable as follows:
Consider the operator in Chern-Simons theory
\eqn\gen{
Z(U,V)=\exp\bigl[ \sum_{n=1}^\infty {1 \over n} {\rm Tr}\, U^n\, 
{\rm Tr }\, V^n\bigr],}
where $U$ is the holonomy of the $U(N)$ Chern-Simons gauge field 
around a given knot, and $V$ is a $U(M)$ matrix. It is convenient 
to expand the exponential in \gen, and the result can be written as 
follows. Let $\vec k$ be a vector with an infinite number of entries, 
almost all zero, and whose nonzero entries are positive integers. 
Given such a vector, we define, as in \quantis:
\eqn\long{
\ell=\sum j \,k_j,\,\,\,\,\,\ |\vec k| =\sum k_j.} 
We can associate to any vector $\vec k$ a conjugacy class $C(\vec k)$ of the  
permutation group  $S_\ell$. This class has $k_1$ cycles of 
length 1, $k_2$ cycles of length 2, and so on. The number of elements 
of the permutation group in such a 
class is given by $\ell!/z_{\vec k}$, where
\eqn\conjug{
z_{\vec k}= \prod_j  k_j! \, j^{k_j}.}
We now introduce the following operators, 
labeled by the vectors $\vec k$:
\eqn\basis{
\Upsilon_{\vec k}(U)= \prod_{j=1}^\infty \Big( {\rm Tr}\,U^j \Big)^{k_j}.}
It is easy to see that:
\eqn\rewr{
Z(U,V)=1+\sum_{\vec k} {1 \over z_{\vec k}}
 \Upsilon_{\vec k} (U) \Upsilon_{\vec k}(V),}   
since we are assuming $\ell>0$. The basis of operators 
\basis, labeled by conjugacy classes of the permutation 
group, is related to the operators labeled by representations 
$R$ of $U(N)$ by Frobenius formula, 
\eqn\frobfor{
\Upsilon_{\vec k}(U) =\sum_R \chi_R (C(\vec k)) {\rm Tr}_R \,U,}
where $\chi_R (C(\vec k))$ is the character of 
the conjugacy class $C(\vec k)$ in the representation labeled by the 
Young tableau of $R$. In particular, one has
\eqn\wk{
W_{\vec k}\equiv \langle \Upsilon_{\vec k}(U) 
\rangle =
\sum_R \chi_R (C(\vec k)) W_R.}
Using \frobfor, one can also write \gen\ as
\eqn\genr{
Z(U,V)= 1 + \sum_R {\rm Tr}_R \, U \, {\rm Tr}_R \,V,}
where the sum over $R$ starts with the fundamental representation.

The main statement of \ov\ is 
\eqn\wilop{\langle Z(U,V)\rangle ={\rm exp}(-F(V))=\exp\biggl(\sum_{n=1}^\infty
\sum_R {1\over n} f_R(q^n, \lambda ^n) {\rm Tr}_R V^n \biggr).}
To check this prediction, given that the Chern-Simons
amplitudes are computable, one can find $f_R$ and see 
if it has the predicted integrality properties. In fact
one can show that
this equation defines the $f_R (q, \lambda)$ 
in terms of $W_R$. The explicit equation is \newpoint\lm\
\eqn\explinver{
\eqalign{
f_R (q, \lambda)=& \sum_{d, m=1}^{\infty} (-1)^{m-1} {\mu (d) \over d m} 
\sum_{\vec k_1, \cdots, \vec k_m} \sum_{R_1, \cdots, R_m} \chi_R ( 
C((\sum_{j=1}^l \vec k_j)_d)) \cr
&\times \prod_{j=1}^m { \chi_{R_j}(C(\vec k_j)) \over z_{\vec k_j}} 
W_{R_j}(q^d, \lambda^d),}} 
where $\vec k_d$ is defined as follows: $(\vec k_d)_{di}=k_i$ and has zero 
entries for the other components. Therefore, if $\vec k= (k_1, k_2, \cdots)$, 
then
$$
\vec k_d =(0, \cdots, 0,k_1,0,\cdots, 0, k_2, 0,\cdots)$$
where $k_1$ is in the $d$-th entry, $k_2$ is in the $2d$-th entry, and so 
on. The sum over $\vec k_1, \cdots, \vec k_m$ is over all vectors with 
$|\vec k_j| >0$. In \explinver, $\mu (d)$ denotes the Moebius function. 
Recall that the Moebius function is defined as follows: if $d$ has the 
the prime decomposition $d=\prod_{i=1}^a p_i^{m_i}$, then $\mu(d)=0$ 
if any of the $m_i$ is greater than one. 
If all $m_i=1$ ({\it i.e.} $d$ is square-free) 
then $\mu(d)=(-1)^a$.
Some examples of \explinver\ are 
\eqn\examples{  
\eqalign{
f_{\tableau{1}}(q,\lambda)=&W_{\tableau{1}}(q, \lambda), 
\cr
f_{\tableau{2}}(q,\lambda)=&W_{\tableau{2}}(q,\lambda)
-{1\over 2}\bigl( W_{\tableau{1}}(q,\lambda)^2+ 
W_{\tableau{1}}(q^2,\lambda^2) 
\bigr),\cr
f_{\tableau{1 1}}(q, \lambda)=&W_{\tableau{1 1}}(q,\lambda)
-{1\over 2}\bigl( W_{\tableau{1}}(q,\lambda)^2-
 W_{\tableau{1}}(q^2,\lambda^2) 
\bigr).\cr }}
These results have been generalized to links in \lmv\newpoint. 
Moreover it has been checked that they satisfy the integrality constraint
predicted in \fr\ in many examples.
In the above 
equations, we have assumed that all the knot invariants have been 
computed in the standard framing. 

Notice that the logarithm of $\langle Z(U,V) \rangle$ is the generating 
function of connected vevs $W_{\vec k}^{(c)}$, 
\eqn\logz{
\log \langle Z(U,V) \rangle = \sum_{\vec k} {1 \over z_{\vec k}}
W_{\vec k}^{(c)} \Upsilon_{\vec k}(V).} Therefore, the free energies of open 
strings are given by:
\eqn\freelog{
i^{|\vec k|} \sum_{g=0}^{\infty} F_{g, \vec k} (t) g_s ^{2g-2 + |\vec k|} 
=-{1 \over \prod_j j^{k_j}} W_{\vec k}^{(c)}.}     

\subsec{Framing dependence of the integer invariants}

Since the Chern-Simons vevs change under a change of framing in the 
way prescribed by \unframing, it is natural to ask what is 
the effect of this change on the invariants 
${\widehat N}_{R,g,Q}$.  In fact as noted in \akv\ for {\it any}
choice of framing we should get integer values for ${\widehat N}_{R,g,Q}$
related to the BPS degeneracies of domain walls in a geometry with
different IR behavior. 
 Notice that, from the point of view of Chern-Simons 
theory, the fact that ${\widehat N}_{R,g,Q}$ are integers in the 
standard framing is already highly nontrivial, and it provides 
one of the major evidences we have for the duality advocated in \gv .  
The integrality predictions for {\it any} $p$ are even more 
surprising. 

In principle, one would think that the change of 
framing for the integer invariants can be determined by 
using the change of framing \unframing, and then by plugging the 
new vevs (which depend on $p$) in \explinver. This should give 
$p$-dependent functions $f_R(q, \lambda)$ from which one can 
extract ${\widehat N}_{R,g,Q}(p)$. It turns out that 
there is a subtlety here \akv:  we should expand the partition 
function $F$ in terms of {\it flat} coordinates
which in this case does depend on the choice of framing
in a simple form.
In fact, the appropriate redefinition of $V$ can be 
read from the results in \av: the $V$ corresponds to 
${\rm e}^{\hat u}$ of \akv. Since under a change of framing 
one has to redefine $\hat u \rightarrow \hat u_p= \hat u + ip\pi$, the natural 
redefinition of $V$ turns out to be
\eqn\redefv{
V\rightarrow (-1)^p V,}
and this means that
\eqn\tracered{
{\rm Tr}_R \, V\rightarrow (-1)^{\ell p} \, {\rm Tr}_R \, V,}
where $\ell$ is the number of boxes in $R$. By looking at 
the generating functional \genr, it is clear that the effect 
of the redefinition is to change the sign of $W_R$ by $(-1)^{\ell p}$. 
This gives a relative sign $f_R$ which is crucial for integrality. 
To compare to topological string theory, it is also useful to reabsorb 
the $\lambda^{p\over 2}$ factor of \unframing\ in $V$ (notice 
that the $\lambda^{p\ell \over 2}$ gives a global factor in $f_R$, since 
all the ``lower order" terms in \explinver\ change by the same factor). 

The main conclusion is that, for a framing labeled by the integer $p$, 
the integer invariants ${\widehat N}_{R,g,Q}(p)$ are obtained 
from \explinver\ but with the vevs 
\eqn\framefin{
W^{(p)}_R (q, \lambda) = (-1)^{\ell p} q^{ {1\over 2} p \kappa_R} 
W_R (q, \lambda),}
where $\kappa_R$ is defined in \kapar. One has, for example, 
\eqn\pexamples{  
\eqalign{
f^{(p)}_{\tableau{1}}(q,\lambda)=&(-1)^pW_{\tableau{1}}(q, \lambda), 
\cr
f^{(p)}_{\tableau{2}}(q,\lambda)=&q^p W_{\tableau{2}}(q,\lambda)
-{1\over 2}\bigl( W_{\tableau{1}}(q,\lambda)^2+ (-1)^p 
W_{\tableau{1}}(q^2,\lambda^2) 
\bigr),\cr
f^{(p)}_{\tableau{1 1}}(q, \lambda)=&q^{-p}W_{\tableau{1 1}}(q,\lambda)
-{1\over 2}\bigl( W_{\tableau{1}}(q,\lambda)^2-(-1)^p
 W_{\tableau{1}}(q^2,\lambda^2) 
\bigr),\cr }}
and so on. Notice that the right framing factor in order to match 
the topological string theory prediction is \explcas, and not \explcasun. 
This is yet another indication that the duality of \gv\ involves the 
$U(N)$ gauge group, not the $SU(N)$ group. The extension to links is 
straightforward: one has just to include the factor in \framefin\ for each 
component.   

\subsec{Examples} 

Using now the explicit expressions for the $U(N)$ Wilson loops, 
together with \explinver, \fr\ and \framefin, one can check that the 
invariants ${\widehat N}_{R,g,Q}$ are in fact integer for any $p$.
We present in the following tables some of the results for 
the unknot and for the trefoil knot (some of the invariants of the 
trefoil knot were computed in \lmv\ in the standard framing). One sees 
that, for given $R$ and $p$, there is only a finite number of $g$, $Q$ 
for which ${\widehat N}_{R,g,Q} \not= 0$. As we increase $p$ 
in absolute value, the $g$'s which have a nonzero invariant also increase.  

\bigskip

{\vbox{\ninepoint{
$$
\vbox{\offinterlineskip\tabskip=0pt
\halign{\strut
\vrule#
&
&\hfil ~$#$
&\hfil ~$#$
&\hfil ~$#$
&\vrule#\cr
\noalign{\hrule}
&Q
&g=0
&g=1
&
\cr
\noalign{\hrule}
&-1
&{1\over 8}(1 -(-1)^p -4p + 2p^2)
&{1\over 96}(-3+3(-1)^p +8p + 4p^2 -8p^3 + 2p^4)
&\cr
\noalign{\hrule}
&0
&-{1\over 2} p (p-1)   
&-{1\over 24}p(p+1)(p-1) (p-2) 
&
\cr
\noalign{\hrule}
&1
&{1\over 8}(-1+(-1)^p + 2p^2) 
&{1\over 96}(3 -3(-1)^p -8p^2 + 2p^4)
&\cr 
\noalign{\hrule}}\hrule}$$}
\vskip - 7 mm
\centerline{{\bf Table 1:} The integers ${\widehat N}_{\tableau{2},g,Q}(p)$
for 
the framed unknot.}
\vskip7pt}
\noindent

\smallskip

{\vbox{\ninepoint{
$$
\vbox{\offinterlineskip\tabskip=0pt
\halign{\strut
\vrule#
&
&\hfil ~$#$
&\hfil ~$#$
&\vrule#\cr
\noalign{\hrule}
&Q
&g=2
&
\cr
\noalign{\hrule}
&-1
&{1\over 5760}(45 -45 (-1)^p -96p -104p^2 + 120 p^3 + 10p^4 -24 p^5 + 4p^6)
&\cr\noalign{\hrule}
&0
&-{1\over 720} (p-3)(p-2)(p-1)p (p+1) (p+2)   
&\cr\noalign{\hrule}
&1
&{1\over 5760}(-45 +45 (-1)^p + 136 p^2 -50 p^4 + 4p^6)
&\cr
\noalign{\hrule}}\hrule}$$}
\vskip - 7 mm
\centerline{{\bf Table 2:} The integers ${\widehat N}_{\tableau{2},g,Q}(p)$
for 
the framed unknot (continuation).}
\vskip7pt}
\noindent

\smallskip

{\vbox{\ninepoint{
$$
\vbox{\offinterlineskip\tabskip=0pt
\halign{\strut
\vrule#
&
&\hfil ~$#$
&\hfil ~$#$
&\hfil ~$#$
&\vrule#\cr
\noalign{\hrule}
&Q
&g=0
&g=1
&
\cr
\noalign{\hrule}
&-1
&{1\over 8}(-1 +(-1)^p + 2p^2)
&{1\over 96}(3 -3(-1)^p -8p^2 + 2p^4)
&\cr\noalign{\hrule}
&0
&-{1\over 2} p (p+1)   
&-{1\over 24}p(p+1)(p-1) (p+2) 
&\cr\noalign{\hrule}
&1
&{1\over 8}(1-(-1)^p + 4p +2 p^2) 
&{1\over 96}(-3 +3(-1)^p -8p + 4p^2 + 8 p^3 + 2p^4 ) 
&\cr 
\noalign{\hrule}}\hrule}$$}
\vskip - 7 mm
\centerline{{\bf Table 3:} The integers 
${\widehat N}_{\tableau{1 1},g,Q}(p)$
for the framed unknot.}
\vskip7pt}
\noindent
\smallskip

\smallskip

{\vbox{\ninepoint{
$$
\vbox{\offinterlineskip\tabskip=0pt
\halign{\strut
\vrule#
&
&\hfil ~$#$
&\hfil ~$#$
&\vrule#\cr
\noalign{\hrule}
&Q
&g=2
&
\cr
\noalign{\hrule}
&-1
&{1\over 5760}(-45 +45 (-1)^p + 136 p^2 -50 p^4 + 4p^6)
&\cr\noalign{\hrule}
&0
&-{1\over 720} (p+3)(p+2)(p+1)p (p-1) (p-2)   
&\cr\noalign{\hrule}
&1
&{1\over 5760}(45 -45 (-1)^p + 
96p -104p^2 - 120 p^3 + 10p^4 +24 p^5 + 4p^6) 
&\cr
\noalign{\hrule}}\hrule}$$}
\vskip - 7 mm
\centerline{{\bf Table 4:} The integers ${\widehat N}_{\tableau{1 1},g,Q}(p)$
for the framed unknot (continuation).}
\vskip7pt}
\noindent

\smallskip

{\vbox{\ninepoint{
$$
\vbox{\offinterlineskip\tabskip=0pt
\halign{\strut
\vrule#
&
&\hfil ~$#$
&\hfil ~$#$
&\hfil ~$#$
&\vrule#\cr
\noalign{\hrule}
&Q
&g=0
&g=1
&
\cr
\noalign{\hrule}
&-3/2
&{1\over 6}(-1)^p p (p-2)(p-1)^2
&{1\over 36}(-1)^p p(p-2)(p-1)^2 (-3 -4p + 2p^2)
&\cr\noalign{\hrule}
&-1/2
&-{1\over 6}(-1)^p p (p-1)(1-5p +3p^2)  
&-{1 \over 24}(-1)^p p(p-2)(p-1)(1-3p -4p^2 + 4p^3)
&\cr\noalign{\hrule}
&1/2
& {1\over 6}(-1)^p p (p-1)(-1-p +3p^2)
&{1 \over 24}(-1)^p p(p+1)(p-1)(2+p -8p^2 + 4p^3)
&\cr\noalign{\hrule}
&3/2
&-{1\over 6}(-1)^p p^2 (p-1)(p+1)
&-{1 \over 36} (-1)^p p^2 (p-1)(p+1)(-5 + 2p^2)
&\cr
\noalign{\hrule}}\hrule}$$}
\vskip - 7 mm
\centerline{{\bf Table 5:} The integers 
${\widehat N}_{\tableau{3},g,Q}(p)$
for 
the framed unknot.}
\vskip7pt}
\noindent

\smallskip

{\vbox{\ninepoint{
$$
\vbox{\offinterlineskip\tabskip=0pt
\halign{\strut
\vrule#
&
&\hfil ~$#$
&\hfil ~$#$
&\hfil ~$#$
&\vrule#\cr
\noalign{\hrule}
&Q
&g=0
&g=1
&
\cr
\noalign{\hrule}
&-3/2
&{1\over 6}(-1)^p p (p-1)(-1-2p + 2p^2)
&{1\over 72}(-1)^p p(p-2)(p-1)(p+1) (-3 -8p + 8p^2)
&\cr\noalign{\hrule}
&-1/2
&-{1\over 6}(-1)^p p (p-1)(-1+2p +6p^2)  
&-{1 \over 24}(-1)^p p(p-1)(p+1)(2-5p -8p^2 + 8p^3)
&\cr\noalign{\hrule}
&1/2
& {1\over 6}(-1)^p p (p+1)(-1-2p +6p^2)
&{1 \over 24}(-1)^p p(p+1)(p-1)(-2-5p+ 8p^2 + 8p^3)
&\cr\noalign{\hrule}
&3/2
&-{1\over 6}(-1)^p p (p+1)(-1+2p + 2p^2)
&-{1\over 72}(-1)^p p(p+2)(p-1)(p+1) (-3 +8p + 8p^2)
&\cr
\noalign{\hrule}}\hrule}$$}
\vskip - 7 mm
\centerline{{\bf Table 6:} The integers 
${\widehat N}_{\tableau{2 1},g,Q}(p)$
for 
the framed unknot.}
\vskip7pt}
\noindent
\smallskip

\smallskip

\smallskip

{\vbox{\ninepoint{
$$
\vbox{\offinterlineskip\tabskip=0pt
\halign{\strut
\vrule#
&
&\hfil ~$#$
&\hfil ~$#$
&\hfil ~$#$
&\vrule#\cr
\noalign{\hrule}
&Q
&g=0
&g=1
&
\cr
\noalign{\hrule}
&-3/2
&{1\over 6}(-1)^p p^2 (p-1)(p+1)
&{1 \over 36} (-1)^p p^2 (p-1)(p+1)(-5 + 2p^2)
&\cr\noalign{\hrule}
&-1/2
& -{1\over 6}(-1)^p p (p+1)(-1+p +3p^2) 
&-{1 \over 24}(-1)^p p(p+1)(p-1)(-2+p +8p^2 + 4p^3)
&\cr\noalign{\hrule}
&1/2
& {1\over 6}(-1)^p p (p+1)(1+5p +3p^2) 
&{1 \over 24}(-1)^p p(p+2)(p+1)(-1-3p +4p^2 + 4p^3)
&\cr\noalign{\hrule}
&3/2
&-{1\over 6}(-1)^p p (p+2)(p+1)^2
&-{1\over 36}(-1)^p p(p+2)(p+1)^2 (-3 +4p + 2 p^2)
&\cr 
\noalign{\hrule}}\hrule}$$}
\vskip - 7 mm
\centerline{{\bf Table 7:} The integers 
${\widehat N}_{\tableau{1 1 1},g,Q}(p)$
for 
the framed unknot.}
\vskip7pt}
\noindent

The above integer invariants for the unknot satisfy the relation
\eqn\refsym{
{\widehat N}_{R,g, Q}(-p) =(-1)^{\ell} {\widehat N}_{R^t,g,-Q} (p),}
where $R^t$ denotes the representation whose Young tableau is 
transposed to the Young tableau of $R$.
This symmetry is easy to explain. Given a knot ${\cal K}$ and its 
mirror image ${\cal K}^*$, the vevs of the corresponding Wilson loops are 
related as follows (see for example \guada):
\eqn\mirror{
W_R(q,\lambda)({\cal K}^*)=W_R(q^{-1},\lambda^{-1})({\cal K}).}
Using \fr, it is easy to see that \mirror\ implies the following 
relation for the integer invariants:
\eqn\mirn{
{\widehat N}_{R,g,Q}({\cal K}^*)=
(-1)^{\ell}{\widehat N}_{R^t,g,-Q}({\cal K}).}
Since the mirror image of the unknot with 
framing $p$ is the unknot with framing $-p$, \refsym\ follows from \mirn.

\smallskip

{\vbox{\ninepoint{
$$
\vbox{\offinterlineskip\tabskip=0pt
\halign{\strut
\vrule#
&
&\hfil ~$#$
&\hfil ~$#$
&\hfil ~$#$
&\vrule#\cr
\noalign{\hrule}
&Q
&g=0
&g=1
&
\cr
\noalign{\hrule}
&1
&{1\over 4}(9-(-1)^p + 2p + 4p^2)
&{1\over 48}(69 - 21 (-1)^p - 4p +98p^2+4p^3 +4p^4)
&\cr\noalign{\hrule}
&2
&-8-5p-p^2  
&-{1\over 12}(72 + 80 p + 75 p^2+ 10 p^3 +3 p^4)
&\cr\noalign{\hrule}
&3
&{1\over 8}(93 + 3(-1)^p +76 p+ 26p^2) 
&{1\over 96} (921 + 39 (-1)^p + 1288 p + 724 p^2 + 152 p^3 + 26p^4)
&\cr\noalign{\hrule}
&4
&-{1\over 2} (16 + 13p + 3p^2) 
&-{1\over 24}(2+p)(72 + 65 p + 20 p^2 + 3p^3)
&\cr\noalign{\hrule}
&5
&{1\over 8}(17 -(-1)^p + 12p + 2p^2)
&{1\over 96}(93 + 3 (-1)^p  + 168 p + 100 p^2 +24p^3+  2p^4)
&\cr 
\noalign{\hrule}}\hrule}$$}
\vskip - 7 mm
\centerline{{\bf Table 8:} The integers ${\widehat N}_{\tableau{2},g,Q}(p)$
for 
the framed trefoil knot.}
\vskip7pt}
\noindent
\smallskip

\smallskip

{\vbox{\ninepoint{
$$
\vbox{\offinterlineskip\tabskip=0pt
\halign{\strut
\vrule#
&
&\hfil ~$#$
&\hfil ~$#$
&\hfil ~$#$
&\vrule#\cr
\noalign{\hrule}
&Q
&g=0
&g=1
&
\cr
\noalign{\hrule}
&1
&{1\over 4}(15+(-1)^p + 10p + 4p^2)
&{1\over 48}(171 + 21 (-1)^p + 220 p +134 p^2+ 20 p^3 +4p^4)
&\cr\noalign{\hrule}
&2
&-16 -11p -3p^2  
&-{1\over 12}(240 + 272 p + 123 p^2+ 22 p^3 +3 p^4)
&\cr\noalign{\hrule}
&3
&{1\over 8}(195 - 3(-1)^p +128 p+ 26p^2) 
&{1\over 96} (3111 - 39 (-1)^p + 3296 p + 1336 p^2 + 256 p^3 + 26 p^4)
&\cr\noalign{\hrule}
&4
&-{1\over 2} (32 + 19p + 3p^2) 
&-{1\over 24}(p+3)(160 + 114 p + 29 p^2 + 3p^3)
&\cr\noalign{\hrule}
&5
&{1\over 8}(31 +(-1)^p + 16p + 2p^2)
&{1\over 96}(387 - 3 (-1)^p  + 448 p + 184 p^2 + 32p^3 + 2p^4)
&\cr 
\noalign{\hrule}}\hrule}$$}
\vskip - 7 mm
\centerline{{\bf Table 9:} The integers ${\widehat N}_{\tableau{1 1},g,Q}(p)$
for 
the framed trefoil knot.}
\vskip7pt}
\noindent
\smallskip

\newsec{Comparison with the direct A-model computation}

In the case of the unknot with an arbitrary framing given by $p$, 
one can find a rather explicit expression 
for the connected vevs. The dual geometry for the unknot is known \ov, and 
some explicit computations of open 
Gromov-Witten invariants for this geometry have been done using 
localization techniques \kl\ls\zas. In particular, Katz and Liu \kl\ were able 
to give an explicit expression for some of these invariants in an arbitrary 
framing, and therefore comparing the 
Chern-Simons answer with their computation 
gives a very powerful check of the duality. Some checks (for genus $0$) have 
been done already in \kl\ in comparison with disk amplitudes of large $N$
Chern-Simons dual
\akv. An interesting corollary of the comparison, as we will discuss,
is that 
all the correlation functions of two-dimensional topological gravity, 
and all Hodge integrals involving up to three $\lambda$ classes
(Chern classes of the Hodge bundle) can be computed 
from the quantum dimensions of a Wess-Zumino-Witten model! In particular, the 
multicovering formula \multicov\ predicts that some combinations of Hodge 
integrals are integers. 

The computation of the connected vevs for the unknot in an arbitrary 
framing is in principle straightforward. A well-known result in Chern-Simons 
theory \cs\ is that the vev of the unknot in the representation $R$ (in the 
standard framing) is given by:
\eqn\quantumd{
W_R = {S_{\rho, \rho + \Lambda} \over S_{\rho,\rho}},}
where $S_{ij}$ are the entries of the $S$-matrix of the $SU(N)$ WZW model, 
$\Lambda$ is the highest weight associated to the representation $R$, and 
$\rho$ is the Weyl vector that represents the vacuum state. 
The right-hand side of \quantumd\ can be written in terms of a character of 
$SU(N)$, and it is also called the quantum dimension of $R$:
\eqn\quantumsi{
{\rm dim}_q R = {\rm ch}_{\Lambda} \biggl[ -{ 2\pi i \over k+N} \rho\biggr].}
The quantum dimension can be explicitly written in terms of $q$-numbers 
as follows. Define:
\eqn\qnumbers{
[x]=q^{x\over 2} -q^{-{x\over 2}},\,\,\,\ [x]_{\lambda} = \lambda^{1\over 2} 
q^{x\over 2} -\lambda^{-{1\over 2}} q^{-{x\over 2}}.}
If $R$ has a Young tableau with $c_R$ rows of lengths $l_i$, 
$i=1, \cdots, c_R$, 
then the quantum dimension can be explicitly written as:
\eqn\expf{
{\rm dim}_q R = \prod_{1\le i < j \le c_R} {[l_i -l_j +j-i] 
\over [j-i]} \prod_{i=1}^{c_R} { \prod_{v=-i+1}^{l_i -i} [v]_{\lambda} 
\over \prod_{v=1}^{l_i} [v-i + c_R]}.}
Now we have all the ingredients to compute $F_{g, \vec k} (t)$ in an 
arbitrary framing given by $p$. According to \freelog, the generating 
functional for $F_{g, \vec k} (t)$ is determined by the connected vev 
$W_{\vec k}^{(c)}$. Therefore, we just have to correct the $W_R$ given in \expf\ 
by the framing factor, compute the $W_{\vec k}$ with  
Frobenius formula, and then extract the connected piece by using:
\eqn\connew{
W_{\vec k}^{(c)} =\sum_{n\ge 1} {(-1)^{n-1} \over n} 
\sum_{\vec k_1, \cdots, \vec k_n} \delta_{\sum_{i=1}^n \vec k_i, \vec k} 
\prod_{i=1}^n {W_{\vec k_i} \over z_{\vec k_i}}.}
In this equation, the second sum is over $n$ vectors 
$\vec k_1, \cdots, \vec k_n$ 
such that $\sum_{i=1}^n \vec k_i =\vec k$ (as indicated by 
the Kronecker delta), 
and therefore the right hand side of \connew\ involves a 
finite number of terms. The generating functional for the open Gromov-Witten 
invariants is then explicitly given by
\eqn\moreor{
\eqalign{
& \sum_{Q}\sum_{g=0}^\infty F_{\vec k, g}^{Q} g_s^{2g-2 + |\vec k|} 
{\rm e}^{Qt} 
={(-1)^{p \ell} \over i^{|\vec k| +\ell} \prod_j j^{k_j}} \sum_{n\ge 1} 
{(-1)^{n} \over n} 
\sum_{\vec k_1, \cdots, \vec k_n} \delta_{\sum_{\sigma=1}^n \vec k_\sigma, \vec k} 
\sum_{R_{\sigma}} \prod_{\sigma =1}^n { \chi_{R_{\sigma}} (C(\vec k_{\sigma}))
\over z_{\vec k_{\sigma}} }\cr & \cdot  
{\rm e}^{i p \kappa_{R_{\sigma}}g_s/2} 
\prod_{ 1\le i < j \le c_{R_{\sigma}}} 
{\sin \Bigl[ (l^{\sigma}_i -l^{\sigma}_j +j-i) g_s/2\Bigr] 
\over \sin \Bigl[ (j-i) g_s /2\Bigr]} 
\prod_{i=1}^{c_{R_{\sigma}}} {\prod_{v=-i+1}^{l_i^{\sigma}-i} 
\bigl( 
{\rm e}^{ {t\over 2} + {iv g_s\over 2}} 
-{\rm e}^{-{t\over 2}-{iv g_s\over 2} }\bigr) 
\over   
\prod_{v=1}^{l^{\sigma}_i} 2 \sin 
\Bigl[ (v-i+c_{R_{\sigma}}) g_s /2\Bigr]}.\cr}} 
The open Gromov-Witten invariants $F_{\vec k, g}^{Q}$ 
have been computed in the A-model by 
Katz and Liu in \kl\ for $Q=\ell/2$, where $\ell=\sum_j jk_j$. 
This corresponds to the leading 
power of ${\rm e}^{Qt}$ in \moreor.  
Their formula is written in terms of the vector $\vec n=(n_1, \cdots, n_h)$ 
and reads:
\eqn\klform{
\eqalign{
F_{\vec n, g}^{\ell/2}=& (-1)^{p\ell}
(p(p+1))^{h-1}\biggl( \prod_{i=1}^h { \prod_{j=1}^{n_i-1} 
(j+n_i p) \over (n_i-1)! } \biggr)
\cr
&\cdot {\rm Res}_{u=0} 
\int_{{\overline M}_{g,h}} 
{c_g (\IE ^{\vee}(u))c_g(\IE ^{\vee} ((-p-1)u)) c_g (\IE ^{\vee} (p u))
u^{2h-4} \over \prod_{i=1}^h (u- n_i \psi_i)}.\cr}}
In this formula, $\overline M_{g,h}$ is the Deligne-Mumford moduli space 
of genus $g$ stable curves with $h$ marked points, and has complex 
dimension $3g-3+h$ (see for example \hm\ for a survey of these 
moduli spaces and their properties). 
$\IE$ is the Hodge bundle over $\overline M_{g,h}$. 
It is a complex vector bundle 
of rank $g$ whose fiber at a point $\Sigma$ is $H^0 (\Sigma, K_{\Sigma})$. 
Its dual is denoted by $\IE ^{\vee} $, and its Chern classes are denoted by:
\eqn\chodge{
\lambda_j =c_j (\IE).}
In \klform, we have written
\eqn\serieshod{
c_g(\IE ^{\vee} (u))= \sum_{i=0}^g c_{g-i} (\IE ^{\vee}) u^i,}
and similarly for the other two factors. The integral in \klform\ also 
involves the $\psi_i$ classes of two-dimensional topological gravity, 
which are defined as follows. We first define the 
line bundle ${\cal L}_i$ over ${\overline M}_{g,h}$ to be the line 
bundle whose fiber over each stable curve $\Sigma$ is the cotangent space 
of $\Sigma$ at $x_i$ (where $x_i$ is the $i$-th marked point). We then have,
\eqn\psiclas{
\psi_i =c_1 ({\cal L}_i),\,\,\,\ i=1, \cdots, h.}
The integrals of the $\psi$ classes can be obtained by the 
results of Witten and Kontsevich on 2d topological gravity 
\wtwodg\k, while the integrals involving 
$\psi$ and $\lambda$ classes (the so-called Hodge integrals) can be in 
principle computed by reducing them to pure $\psi$ integrals \faber. Explicit 
formulae for some Hodge integrals have 
been recently found in \hint. In writing 
\klform, we have taken into account that 
the variable $a$ used in \kl\ corresponds to $-p$ here, and we have 
included the global factor $(-1)^{p\ell}$ 
which is crucial in order to extract 
integer invariants by means of the multicovering formulae.

The Chern-Simons computation \moreor\ gives an explicit 
generating functional 
for all the open Gromov-Witten invariants, including those in \klform\ 
with $Q=\ell/2$. In particular, it is 
in principle possible to compute all the integrals over 
${\overline M}_{g,h}$ that appear in \klform\ from the explicit 
expression \moreor! These Hodge integrals include an arbitrary number 
of $\psi$ classes and up to three $\lambda$ classes. Therefore, all 
correlation functions of two-dimensional topological gravity can in principle 
be extracted from \moreor. It should be noted, however, that some 
of the simple structural properties of \klform\ are not at all obvious from 
\moreor. For example, for $g=0$, $h=1$, \klform\ gives a fairly 
compact expression for the open Gromov-Witten invariant, and the fact 
that this equals the Chern-Simons answer amounts to a rather 
nontrivial combinatorial identity, as it was already observed in \akv. 

Let us compare the two expressions, \klform\ and \moreor, in some 
simple examples with $h=1$. Since for Riemann surfaces with one hole 
the homotopy class of the map is given by the winding number $j$, 
we will denote the invariants by $F_{j,g}^Q$ and $n_{j,g,Q}$ (and 
we are going to take $Q=j/2$ to compare with \akv\kl). 
For $g=1$, one finds:
\eqn\resultone{
F_{j,1}^{j/2} ={(-1)^{pj}\over (j-1)!} \prod_{l=1}^{j-1} (l + jp) \Biggl( 
\biggl(  \int_{ {\overline M}_{1,1}} \lambda_1-j \psi_1 
\biggr) p(p+1) + \int_{ {\overline M}_{1,1}} \lambda_1 \Biggr), }
and for $g=2$, 
\eqn\resultwo{
\eqalign{
F_{j,2}^{j/2} &={(-1)^{pj}\over (j-1)!} \prod_{l=1}^{j-1} (l + jp) \Biggl( 
\biggl(\int_{ {\overline M}_{2,1}} j^2 \psi_1^4 - 
j \psi_1^3 \lambda_1 + \psi_1^2 \lambda_2  \biggr) j^2 p^3 (p+2) \cr
&+ \biggl(\int_{ {\overline M}_{2,1}} j^3 \psi_1^4 - 
2 j^2 \psi_1^3 \lambda_1 - 
\psi_1\lambda_1\lambda_2 + 3j \psi_1^2\lambda_2  \biggr) 
jp^2 \cr
& +  \biggl(\int_{ {\overline M}_{2,1}} -j^2 \psi_1^3 \lambda_1 - 
\psi_1\lambda_1 \lambda_2 + 2j \psi_1^2\lambda_2  \biggr) 
jp  + j^2 \int_{ {\overline M}_{2,1}} \psi_1^2\lambda_2\Biggr) .\cr}}
To obtain this expression, we have used the Mumford relations $\lambda_2^2=0$ 
and $\lambda_1^2 =2\lambda_2$ \mumford. 
All the integrals involved here can 
be extracted from the generating functional 
\moreor, computed up to order $g_s^4$, and for 
two values of $j$, say $j=1,2$. These are 
easily computed to be:
\eqn\computa{
\eqalign{
iW_1^{(c)} (g_s) = &{(-1)^p \over g_s}\biggl( 1 + {1 \over 24} g_s^2 + 
{7 \over 5760} g_s^4+ {\cal O}(g_s^6)\biggr),\cr 
{i\over 2} W_2^{(c)}(g_s) = & { 1+ 2p 
\over g_s}  \biggl( {1 \over 4} 
- {1 \over 24} (p^2 +p - 1) g_s^2 \cr & \,\,\  + 
{1 \over 1440}(7-11p -8p^2 + 6p^3 + 3p^4)g_s^4 
+ {\cal O}(g_s^6)\biggr).\cr}}
After some simple algebra, one finds for $g=1$:
\eqn\genone{
\int_{ {\overline M}_{1,1}}\psi_1= 
\int_{ {\overline M}_{1,1}}\lambda_1={1\over 24}
}and for $g=2$
$$
\int_{ {\overline M}_{2,1}}\psi_1^4={1\over 1152},\,\,\,\ 
\int_{ {\overline M}_{2,1}}\psi_1^3\lambda_1 ={1 \over 480}, $$
\eqn\someal{
\int_{ {\overline M}_{2,1}}\psi_1^2\lambda_2 ={7\over 5760},\,\,\,\
\int_{ {\overline M}_{2,1}}\psi_1\lambda_1\lambda_2 ={1 \over 2880},}
in agreement with known results (see for example \hint). 

In general, for arbitrary $h$, the coefficient of the leading power of 
$p$ is a sum of Hodge integrals which includes
\eqn\twodg{
\sum_{k_1, \cdots, k_h} n_1^{k_1} \cdots n_h^{k_h} \int_{{\overline M}_{g,h}} 
\psi_1^{k_1} \cdots \psi_h^{k_h},}
so in principle one can extract the correlation functions of 2d topological 
gravity from \moreor. This in turn suggests 
that the Kontsevich matrix integral \k , which
can be viewed as a large $N$ duality for a 0-dimensional
gauge theory,
may be obtained in an appropriate limit from the large $N$ duality
conjecture of \gv\ 
for three dimensional Chern-Simons gauge theory with an insertion 
of observables like \gen.  This would be very interesting to directly
establish.

One can obtain integer invariants from the Gromov-Witten invariants by 
using the multicovering formulae. The most basic integer invariants 
are the integers ${\widehat N}_{R,g,Q} (p)$, and one can compute the 
integers $n_{\vec k, g,Q}$ that enter the multicovering 
formula \multicov\ from \nminus. The relevant formulae simplify in some 
particular cases. For example, if we want to compute 
$n_{j,g,Q}$ (corresponding 
to one hole with wrapping number $j$), then the generating 
functional
\eqn\genfunct{
f_{j}(q, \lambda)= (q^{-{j\over 2}} -q^{j\over 2})\sum_{g,Q} 
n_{j,g,Q} (q^{-{1\over 2}} - q^{1\over 2})^{2g-2} \lambda^Q}
is given by
\eqn\genfunct{
f_j (q,\lambda)= \sum_{d|j} \mu(d) W_{j/d}^{(c)}(q^d, \lambda^d),}
and we recall that $\mu(d)$ is the Moebius function. 
This expression can be derived from \fr\ and \explinver\ 
(see \lmv\ for more details). In the case $h=1$, the integer invariants 
that correspond to the open Gromov-Witten invariants $F_{j,g}^{j/2}$ computed 
by Katz and Liu are $n_{j,g,j}$. The relation between them is 
precisely the one written in \expl\ for lower genera. In the following 
tables we list some of these integer invariants. For $g=0$, 
these results were obtained in \akv\ in the context of the B-model.   
                       
\smallskip

{\vbox{\ninepoint{
$$
\vbox{\offinterlineskip\tabskip=0pt
\halign{\strut
\vrule#
&
&\hfil ~$#$
&\hfil ~$#$
&\hfil ~$#$
&\vrule#\cr
\noalign{\hrule}
&j
&g=0
&g=1
&
\cr
\noalign{\hrule}
&1
&-(-1)^p
&0
&\cr\noalign{\hrule}
&2
&-{1\over 4}(2p +1 -(-1)^p)  
&-{1\over 48}(-3+3(-1)^p - 4p + 6p^2 + 4p^3)
&\cr\noalign{\hrule}
&3
&-{1\over 2}(-1)^p p(p+1)
&-{1\over 8}(-1)^p p(p+1)(-2 + 3p + 3p^2)
&\cr\noalign{\hrule}
&4
&-{1\over 3} p(p+1)(2p+1)
&-{1\over 3}p(p+1)(2p+1)(-1+ 2p + 2p^2)
&\cr\noalign{\hrule}
&5
&-{5\over 24}(-1)^p p(p+1) (2 + 5p + 5p^2)
&-{1\over 144}(-1)^p p(p+1)(-96 -50 p + 575 p^2 + 1250 p^3 
+ 625 p^4) 
&\cr 
\noalign{\hrule}}\hrule}$$}
\vskip - 7 mm
\centerline{{\bf Table 10:} The integers $n_{j,g,j}(p)$
for the framed unknot.}
\vskip7pt}
\noindent
\smallskip

\smallskip

{\vbox{\ninepoint{
$$
\vbox{\offinterlineskip\tabskip=0pt
\halign{\strut
\vrule#
&
&\hfil ~$#$
&\hfil ~$#$
&\vrule#\cr
\noalign{\hrule}
&j
&g=2
&
\cr
\noalign{\hrule}
&1
&0
&\cr\noalign{\hrule}
&2
&-{1\over 960}(15-15(-1)^p +16p -40 p^2 -20 p^3 + 10p^4 + 4p^5 )  
&\cr\noalign{\hrule}
&3
&-{1 \over 80}(-1)^p p(p-1)(p+1)(2+p)(3p-1)(3p+4)
&\cr\noalign{\hrule}
&4
&-{1\over 90} p (p+1)(1+2p) (21 -77 p -25p^2 + 104 p^3 
+ 52 p^4)
&\cr\noalign{\hrule}
&5
&-{1\over 384}(-1)^p p(p+1)(-4 + 5p + 5p^2)(-68 + 625 p^2 
+ 1250 p^3 + 625 p^4)
&\cr 
\noalign{\hrule}}\hrule}$$}
\vskip - 7 mm
\centerline{{\bf Table 11:} The integers $n_{j,g,j}(p)$
for the framed unknot (continuation).}
\vskip7pt}
\noindent
\smallskip

\newsec{The framed unknot and the B-model disk amplitude}

The open Gromov-Witten invariants for $g=0$, $h=1$ (disk amplitudes) 
can also be computed in the B-model by using the techniques of \av\akv. 
In \akv, the mirror geometry to the framed unknot was studied in detail 
in the large volume limit, and this allowed to obtain an explicit expression 
for the invariants $F_{j,g=0}^{j/2}$. In fact, one can extend the computation 
in \akv\ and obtain the explicit expression of $F_{j,g=0}^Q$ for 
$Q=-j/2, \cdots, j/2$, which we will now do, in order to compare
with the results we have obtained from Chern-Simons theory. 

The A-model geometry for the unknot in ${\bf S}^3$ was found in \ov, and 
it is a Lagrangian submanifold in the resolved conifold ${\cal O}(-1) \oplus 
{\cal O}(-1) \rightarrow {\bf P}^1$.   Actually, more precisely it
is the same Lagrangian submanifold discussed in \ov , but in the ``flopped
phase'' of the Lagrangian submanifold
(Phase II, p. 25 of \av )\foot{We are grateful to Mina Aganagic
for discussions leading to this clarification.}.
  This also avoids the problem of having
the ${\bf S}^1$ of the original Lagrangian submanifold getting flopped in the
blow up geometry, which was noticed by the authors of \ov\ (but
not written in the paper).  The mirror geometry for the brane 
is characterized by the Riemann surface,
\eqn\bm{
{\rm e}^{t} {\rm e}^{v+u} + {\rm e}^u + {\rm e}^v +1 =0.}
 The right 
variables to use are, according to the analysis in \akv, $\hat u= u + i\pi$, 
$\hat v = v + i\pi$. This finally gives the equation,
\eqn\bmo{
{\rm e}^{\hat u} + {\rm e}^{\hat v} -
{\rm e}^{t} {\rm e}^{\hat u + \hat v}   =1.}
An arbitrary framing specified by $p$ corresponds to a shift $\hat u 
\rightarrow \hat u + p\hat v$, and the equation to be solved reads:
\eqn\bmod{
xy^p + y -{\rm e}^t x y^{p+1}=1,}
where we have denoted
\eqn\varias{
x={\rm e}^{\hat u},\,\,\,\,  y={\rm e}^{\hat v}.}
The algebraic equation \bmod\ can be solved with the ansatz $y=\sum_{m=0}^
{\infty} a_m x^m$, as in \akv. One gets a recursive relation for the 
coefficients with the following explicit solution:
\eqn\sol{
a_m =\sum_{l=0}^m a_{m,l} \, {\rm e}^{l t},}
where
\eqn\coeffs{
a_{m,l} = {(-1)^{m+l} \over m!} {m \choose l} 
\prod_{j=-l}^{m-l-2} (mp-j).}
The open Gromov-Witten invariants are the coefficients of the superpotential,
\eqn\superpot{
W= \sum_{m=1}^{\infty}\sum_{l=0}^m W_{m,l} \, {\rm e}^{m\hat u + lt},}
which can be obtained from the equation $\hat v =\partial_{\hat u}W$, or 
equivalently $x\partial_x W= \log y$. One then finds,
\eqn\finalres{
W_{m,l}={(-1)^{m+l} \over m\cdot m!} { m \choose l} 
\prod_{j=-l+1}^{m-l-1} (mp-j),\,\,\,\,\,\,\ l=0,\cdots,m.}
The result of \akv\ for the ``almost ${\bf C}^3$'' geometry is a particular 
case of \coeffs\ and \finalres\ when $l=0$. Notice the symmetry
\eqn\wsym{
W_{m,l}(-p)=-W_{m,m-l}(p).} 

We have checked in many cases that the above result agrees with the 
Chern-Simons result for the framed unknot \moreor. More precisely, 
one has that
\eqn\precise{
F_{m,g=0}^Q= (-1)^{p m} W_{m, Q+m/2},\,\,\,\,\ Q=-m/2, \cdots, m/2.}
The multicovering formulae \expl\ predict that, if we write the 
superpotential as
\eqn\supermult{
W=-\sum_{m,l} \sum_{k>0} {n_{m,l} \over k^2} {\rm e}^{k(m\hat u_p + lt)},}
where $\hat u_p =\hat u + \pi i p$, then the $n_{m,l}$ are integers. In fact, 
they are the integer invariants $n_{m,g=0,l-m/2}$ that appear in \expl\ and 
that can be computed from Chern-Simons theory. The case $l=0$ was 
obtained in \akv. For $l>0$ one finds, for example:
$$
n_{2,1}=p,\,\,\,\,\,  n_{2,2}= -{p\over 2} +{(-1)^p-1\over 4},$$
$$
n_{3,1}=-{(-1)^p\over 2}p(3p-1),\,\,\,\,\ n_{3,2}={(-1)^p\over 2}p(3p+1), 
\,\,\,\,\,\, n_{3,3}=-{(-1)^p\over 2} p(p+1),$$
$$
n_{4,1}={p\over 3}(2p-1)(4p-1),\,\,\,\,\, n_{4,2}=-4p^3, \,\,\,\,\, 
n_{4,3}={p\over 3}(2p+1)(4p+1), \,\,\,\,\, n_{4,4}=-{p\over 3}(p+1)(2p+1),$$ 
and so on. For $m=2,3$, one can immediately check that the above expressions 
agree with the invariants \nminus\ obtained from tables 1-7. It follows from 
\wsym\ that  
\eqn\mirns{
n_{m,l}(-p)=-n_{m,m-l}(p),}
which from the Chern-Simons point of view is a consequence of \refsym. 
Notice that, when $p=1$, the integers $n_{m,l}$ are precisely 
the $d_{l,m}$ computed 
in \av\ for phase II of the Lagrangian submanifold in ${\cal O}(-1) 
\oplus {\cal O}(-1) \rightarrow {\bf P}^1$.   

\bigskip

\centerline{\bf Acknowledgements}
\medskip
We would like to thank M. Aganagic, A. Klemm and J.M.F. Labastida 
for valuable discussions.

The research of M.M. is 
supported by DOE grant DE-FG02-96ER40959. 
The research of C.V. is supported in part by NSF grants PHY-9802709
and DMS-0074329.

\listrefs
\bye